\documentclass[twocolumn,showpacs,superscriptaddress,preprintnumbers,amsmath,amssymb]{revtex4}

\usepackage{graphicx}
\usepackage{dcolumn} % Align table columns on decimal point
\usepackage{bm} % bold math
\usepackage[figuresright]{rotating}

\newcommand{\BE}{$B$($E2$)}
\newcommand{\E}{$E$(2$_1^+$)}

\begin{document}

\preprint{submitted to Physical Review Letters}x

\title{Anomalously hindered $E$2 strength 
$B$($E2;2_1^+\rightarrow 0^+$)  in $^{16}$C}

\author{N.~Imai}
\email{imai@rarfaxp.riken.go.jp}
\affiliation{RIKEN, Hirosawa 2-1, Wako, Saitama 351-0198, Japan}
\author{H.J.~Ong}
\affiliation{Department of Physics, University of Tokyo, Hongo 7-3-1,
              Bunkyo, Tokyo 113-0033, Japan }
\author{N.~Aoi}
\affiliation{RIKEN, Hirosawa 2-1, Wako, Saitama 351-0198, Japan}
\author{H.~Sakurai}
\affiliation{Department of Physics, University of Tokyo, Hongo 7-3-1,
              Bunkyo, Tokyo 113-0033, Japan }
\author{K.~Demichi}
\affiliation{Department of Physics, Rikkyo University,
        Nishi-Ikebukuro 3-34-1, Toshima, Tokyo 171-8501, Japan}
\author{H.~Kawasaki}
\affiliation{Department of Physics, Rikkyo University,
        Nishi-Ikebukuro 3-34-1, Toshima, Tokyo 171-8501, Japan}
\author{H.~Baba}
\affiliation{Department of Physics, Rikkyo University,
        Nishi-Ikebukuro 3-34-1, Toshima, Tokyo 171-8501, Japan}
\author{Zs.~Dombr\'adi}
%\affiliation{Institute of Nuclear Research of the Hungarian
%        Academy of Science, H-4001 Debrecen, P.O. Box 51,  Hungary}
\affiliation{ATOMKI, H-4001 Debrecen, P.O. Box 51,  Hungary}
\author{Z.~Elekes}
\thanks{On leave from ATOMKI, Debrecen, Hungary.}
\affiliation{RIKEN, Hirosawa 2-1, Wako, Saitama 351-0198, Japan}
\author{N.~Fukuda}
\affiliation{RIKEN, Hirosawa 2-1, Wako, Saitama 351-0198, Japan}
\author{Zs.~F\"ul\"op}
\affiliation{ATOMKI, H-4001 Debrecen, P.O. Box 51,  Hungary}
%\affiliation{Institute of Nuclear Research of the Hungarian
%        Academy of Science, H-4001 Debrecen, P.O. Box 51,  Hungary}
\author{A.~Gelberg}
\affiliation{Institut f\"ur Kernphysik der Universit\"at zu K\"oln, 
        D-50937 K\"oln, Germany}
\author{T.~Gomi}
\affiliation{Department of Physics, Rikkyo University,
        Nishi-Ikebukuro 3-34-1, Toshima, Tokyo 171-8501, Japan}
\author{H.~Hasegawa}
\affiliation{Department of Physics, Rikkyo University,
        Nishi-Ikebukuro 3-34-1, Toshima, Tokyo 171-8501, Japan}
\author{K.~Ishikawa}
\affiliation{Department of Physics, Tokyo Institute of Technology, 
	Ookayama 2-12-1, Meguro, Tokyo 152-8551, Japan}
\author{H.~Iwasaki}
\affiliation{Department of Physics, University of Tokyo, Hongo 7-3-1,
              Bunkyo, Tokyo 113-0033, Japan }
\author{E.~Kaneko}
\affiliation{Department of Physics, Rikkyo University,
        Nishi-Ikebukuro 3-34-1, Toshima, Tokyo 171-8501, Japan}
\author{S.~Kanno}
\affiliation{Department of Physics, Rikkyo University,
        Nishi-Ikebukuro 3-34-1, Toshima, Tokyo 171-8501, Japan}
\author{T.~Kishida}
\affiliation{RIKEN, Hirosawa 2-1, Wako, Saitama 351-0198, Japan}
\author{Y.~Kondo}
\affiliation{Department of Physics, Tokyo Institute of Technology, 
	Ookayama 2-12-1, Meguro, Tokyo 152-8551, Japan}
\author{T.~Kubo}
\affiliation{RIKEN, Hirosawa 2-1, Wako, Saitama 351-0198, Japan}
\author{K.~Kurita}
\affiliation{Department of Physics, Rikkyo University,
        Nishi-Ikebukuro 3-34-1, Toshima, Tokyo 171-8501, Japan}
\author{S.~Michimasa}
\affiliation{CNS,
              University of Tokyo,
              RIKEN campus, Hirosawa  2-1, Wako, Saitama 351-0198, Japan}
\author{T.~Minemura}
\affiliation{RIKEN, Hirosawa 2-1, Wako, Saitama 351-0198, Japan}
\author{M.~Miura}
\affiliation{Department of Physics, Tokyo Institute of Technology, 
	Ookayama 2-12-1, Meguro, Tokyo 152-8551, Japan}
\author{T.~Motobayashi}
\affiliation{RIKEN, Hirosawa 2-1, Wako, Saitama 351-0198, Japan}

\author{T.~Nakamura}
\affiliation{Department of Physics, Tokyo Institute of Technology, 
	Ookayama 2-12-1, Meguro, Tokyo 152-8551, Japan}
\author{M.~Notani}
\affiliation{CNS,
              University of Tokyo,
              RIKEN campus, Hirosawa 2-1, Wako, Saitama 351-0198, Japan}
\author{T.K.~Ohnishi}
\affiliation{Department of Physics, University of Tokyo, Hongo 7-3-1,
              Bunkyo, Tokyo 113-0033, Japan }
\author{A.~Saito}
\affiliation{Department of Physics, Rikkyo University,
        Nishi-Ikebukuro 3-34-1, Toshima, Tokyo 171-8501, Japan}
\author{S.~Shimoura}
\affiliation{CNS,
              University of Tokyo,
              RIKEN campus, Hirosawa 2-1, Wako, Saitama 351-0198, Japan}
\author{T.~Sugimoto}
\affiliation{Department of Physics, Tokyo Institute of Technology, 
	Ookayama 2-12-1, Meguro, Tokyo 152-8551, Japan}
\author{M.K.~Suzuki}
\affiliation{Department of Physics, University of Tokyo, Hongo 7-3-1,
              Bunkyo, Tokyo 113-0033, Japan }
\author{E.~Takeshita}
\affiliation{Department of Physics, Rikkyo University,
        Nishi-Ikebukuro 3-34-1, Toshima, Tokyo 171-8501, Japan}
\author{S.~Takeuchi}
\affiliation{RIKEN, Hirosawa 2-1, Wako, Saitama 351-0198, Japan}
\author{M.~Tamaki}
\affiliation{CNS,
              University of Tokyo,
              RIKEN campus, Hirosawa 2-1, Wako, Saitama 351-0198, Japan}
\author{K.~Yoneda}
\thanks{Present address: NSCL  %National Superconducting Cyclotron Laboratory,
   Michigan State University, East Lansing, Michigan 48824, USA.}
\affiliation{RIKEN, Hirosawa 2-1, Wako, Saitama 351-0198, Japan}
\author{H.~Watanabe}
\affiliation{RIKEN, Hirosawa 2-1, Wako, Saitama 351-0198, Japan}
\author{M.~Ishihara}
\affiliation{RIKEN, Hirosawa 2-1, Wako, Saitama 351-0198, Japan}

\date{\today}% It is always \today, today,

\begin{abstract}
%%%%%%%%01234567890123456789012345678901234567890123456789
	The electric quadrupole transition from the first 2$^+$ state to the ground 
	0$^+$ state in  $^{16}$C is studied through measurement of the lifetime
	by a recoil shadow method applied to inelastically scattered radioactive 
	$^{16}$C nuclei. 
	The measured lifetime is 75~$\pm$~23~ps, corresponding to a 
	$B$($E2;2_1^+\rightarrow 0^+$) value of 0.63~$\pm$~0.19~$e^2$fm$^4$, or 
	0.26~$\pm$~0.08 Weisskopf units. 
	The transition strength is found to be anomalously small compared to 
	the empirically predicted value.
\end{abstract}

\maketitle
	Quadrupole strengths are fundamental quantities in probing the collective 
	character of nuclei. The enhancement of the electric quadrupole ($E2$) 
	transition strength with respect to that of  single proton excitation may 
	reflect large fluctuation or deformation of nuclear charge~\cite{BM}. 
	One of the important $E2$ transitions in an even-even nucleus is that from 
	the first 2$^+$ (2$^+_1$) state to the ground state (0$^+_{\rm g.s.}$), 
	the reduced transition probability 
%	($B$($E2$;2$^+_1 \rightarrow$ 0$^+_{\rm g.s.}$)) 
	\BE\
	of which has long been a basic observable in the extraction of the magnitude 
	of nuclear deformation or in probing anomalies in the nuclear structure. 
	With recent advances in techniques for supplying intense beams of unstable 
	nuclei, several exotic properties such as magicity loss~\cite{iwasaki,moto,scheit} 
	have been discovered in neutron-rich nuclei through measurements of 
	$E2$ strengths.

	The present Letter reports lifetime measurements of the 2$^+_1$ state of the 
	neutron-rich nucleus  $^{16}$C. The lifetime is inversely proportional to 
	\BE.
%	$B$($E2$;2$^+_1 \rightarrow$ 0$^+_{\rm g.s.}$), which is referred to hereafter 
%	as  \BE. 
	A simple model of a nucleus as a quantum liquid-drop well describes 
	the systematic tendency that \BE\  varies in inverse proportion to the 
	excitation energy $E(2_1^+)$ of the 2$^+_1$ state~\cite{raman2}. 
	For carbon isotopes, when $N$ changes from the magic number 8 to 10, \E\ 
	decreases dramatically from 7.012 to 1.766~MeV~\cite{ToI}. It can then be 
	anticipated that $^{16}$C ($N$~=~10) would
%	 exhibit an appreciable deformation involving 
	have a much larger \BE\ than that of $^{14}$C ($N$~=~8). 
	Unexpectedly, a remarkably small \BE\ was found for $^{16}$C in the 
	present work. The observed value, in Weisskopf units, turns out to be far 
	smaller than any other \BE\ measured on the nuclear chart.

	In the present experiment, a new technique was employed to measure the lifetime 
	of an excited state populated in inverse-kinematics reactions. 
	The technique essentially 
	followed the concept of the recoil shadow method (RSM)~\cite{rsm}, in which 
	the emission point of the de-excitation $\gamma$-ray is located and the 
	$\gamma$-ray intensity is recorded as a function of the flight distance 
	of the de-exciting nucleus. As the flight velocity of the de-exciting nucleus 
	is close to half the velocity of light, the flight distance over 100~ps corresponds 
	to  a macroscopic length of about 1.7~cm. Thus, the present shadow method 
	provides a wide range of applicability, extending to lifetimes of 
%	substantially less than 100~ps. 
	as short as a few tens of ps. 
	In particular, the method is useful for determining the \BE\ 
	value of $Z<10$ nuclei, at which the use of intermediate-energy Coulomb 
	excitation~\cite{moto,scheit} 
%	susceptible to contamination of nuclear excitation.
	may suffer from contamination of nuclear excitation.

	The experiment was performed at the RIKEN accelerator research facility. 
	A secondary $^{16}$C beam was produced through projectile fragmentation of a 
	100-MeV/nucleon $^{18}$O primary beam, separated by the RIPS beam line~\cite{rips}.
	The $^{16}$C beam was directed at a thick, 370-mg/cm$^2$ $^{9}$Be target placed 
	at the exit of the RIPS line for inelastically excitation. 
	Particle identification for the secondary beam was performed event-by-event 
	by means of the time-of-flight (TOF)-$\Delta E$ method using a 1.0~mm-thick 
	plastic scintillation counter (PL) located 180~cm upstream of the target. 
	Two sets of parallel plate avalanche counters (PPACs) were also placed upstream 
	of the target to record the position and angle of the projectile incident upon 
	the target.  The $^{16}$C beam had a typical intensity of 2$\times 10^5$ particles 
	per second and purity of  about 97\%. The average energy across the target was 
	34.6~MeV/nucleon.

	Particles scattered from the target were identified by the $\Delta E$-$E$-TOF 
	method using a plastic scintillator hodoscope located 86~cm downstream of the
	target. The hodoscope, with an active area of 24$\times$24~cm$^2$, consisted of a 
	2.0 mm-thick $\Delta E$ plane and a 5.0 mm thick $E$ plane. 
	The scattering angles were measured by another PPAC placed 25~cm downstream of 
	the target.
 
\begin{figure}
\begin{center}
\includegraphics[width=80mm]{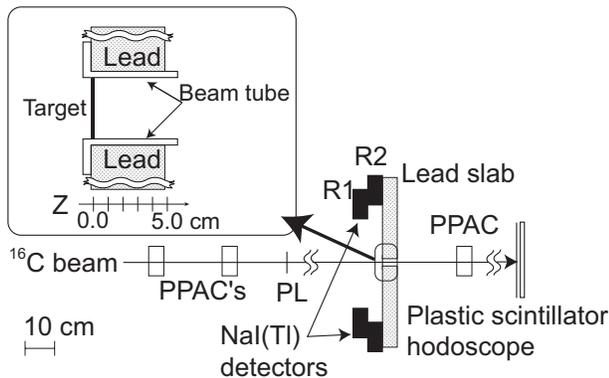}
\end{center}
\caption{Schematic of experimental setup. A beryllium target is surrounded by a 5~cm-thick lead wall, and two cylindrically layers of  NaI(Tl) scintillators (R1 and R2) are placed upstream of the target.}
\label{setup}
\end{figure}

	The experimental setup for $\gamma$-ray detection is shown schematically in 
	Fig.~\ref{setup}. In order to implement the RSM concept, a thick $\gamma$-ray 
	shield was placed around the target. The shield was a 5~cm-thick lead slab with 
	an outer frame of 60$\times$60~cm$^2$ and an inner hole of 6.2 cm~$\phi$. 
	The inner hole surrounded the beam tube housing the $^{9}$Be target. 
	For the sake of later discussion, the $z$-axis is defined as the beam direction, 
        that is close to  the flight direction of the de-exciting nucleus.
%	that is, the flight direction of the de-exciting nucleus. 
	The origin of the $z$-axis, 
	$z=0.0$~cm, was taken at the upstream surface of the lead slab.

	The $\gamma$-rays from the excited $^{16}$C  in-flight were detected 
	by two rings of NaI(Tl) detectors labeled R1 and R2, composed respectively of 
	14 and 18 rectangular NaI(Tl) crystals with volume of $6 \times 6 \times 12$ cm$^3$. 
	These rings circled the beam tube, with crystal centers set at polar 
	angles of 121$^{\circ}$ and 102$^{\circ}$, respectively. The average distance from 
	the target to R1 was 29.2~cm, and 28.7~cm for R2.
     
	The acceptance of the detectors was determined by the geometry of the R1 and R2 
	detectors with respect to the lead slab, and the emission point $z$. 
	Specifically, the $\gamma$-rays emitted between $z=0.0$ and 2.2~cm can reach
	the centers of the NaI crystals of R1 without passing through the lead slab 
	($z=0.0$ and 0.7~cm for R2). Hence, the relative efficiency of R2 with respect 
	to R1 decreased as the emission point was moved further downstream. 
	This difference of geometrical efficiency results in a dependence of
	the  $\gamma$-ray yield ratio (R1/R2) on the lifetime of the decaying state.

\begin{figure}
\begin{center}
\includegraphics[width=80mm]{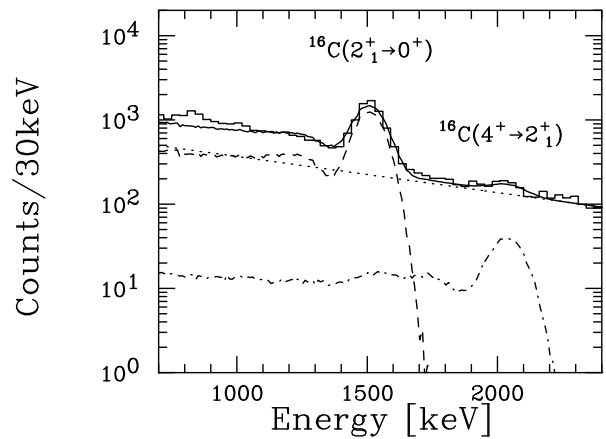}
\end{center}
\caption{
	Energy spectrum of $\gamma$-rays measured  at R1 for target position  $z=0.0$~cm. 
	Dashed line represents the transition of 1766~keV from the excited states 
	of $^{16}$C, and dash-dotted line denotes that for 2376~keV, as simulated using 
	GEANT code~\cite{geant}. The solid curve represents the best fit of the simulated 
	line and an exponential background. The dotted line denotes the quoted background.}
\label{e_spectrum}
\end{figure}

	The measurement of R1/R2 was conducted for two target positions, $z=0.0$ and 
	1.0~cm. Figure~\ref{e_spectrum} shows a Doppler-uncorrected $\gamma$-ray energy 
	spectrum measured by R1 for the target placed at $z=0.0$~cm. The spectrum is 
	dominated by a peak at around 1.5 ~MeV, representing the Doppler-shifted 1766~keV 
	line of the 2$^+_1 \rightarrow$ 0$^+_{\rm g.s.}$ transition in $^{16}$C. 
	A minor peak at around 2.0~MeV is also present, which may correspond to the known 
	4$^+ \rightarrow$ 2$^+_1$ transition in $^{16}$C. As the intensity of the latter 
	transition is only 2\% of the former, the contribution of the minor peak is 
	ignored in subsequent analysis.

	In deducing the yield ratio, a Gaussian shape for the peak and either an exponential 
	or polynomial shape for the background component were fitted to the $\gamma$-ray 
	energy spectrum. The R1/R2 ratios thus obtained for the 1766-keV $\gamma$-rays were 
	1.06~$\pm$~0.03 and 1.70~$\pm$~0.06 for target positions of $z=0.0$ and 1.0~cm, 
	respectively. The errors are statistical errors only. The influence of the choice 
	of functional for the background is negligible. 

	In order to relate the R1/R2 ratio to lifetime, Monte Carlo simulation was performed 
	to evaluate the detection efficiencies of R1 and R2 as a function of the position 
	$z$ of $\gamma$ emission. By folding the efficiency over the flight distance, 
	the R1/R2 ratio can be obtained as a function of mean-life $\tau$. 
	The simulation was performed using GEANT code~\cite{geant}, and involved the shape 
	of the detector and the geometry of the experimental setup. The validity of the 
	simulation was tested by first performing a separate measurement in which a 
	$^{22}$Na standard source emitting 1275-keV $\gamma$~rays was placed at 
	twenty-one positions 
	from $z=0.0$ to 2.0~cm. The R1/R2 ratios thus measured for the different geometries 
	were all reproduced by the simulation within $\pm$~3\%.

\begin{figure}
\begin{center}
\includegraphics[width=80mm]{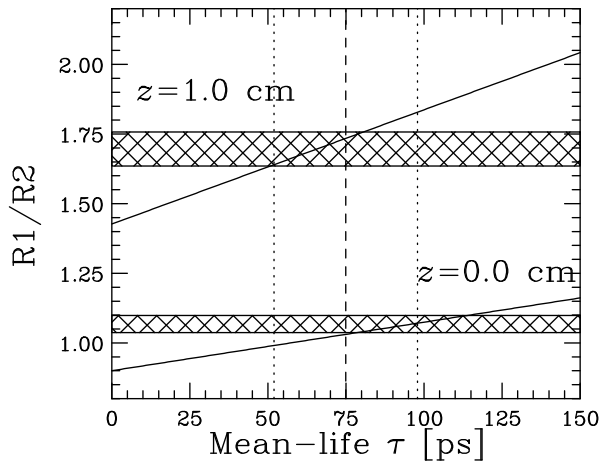}
\end{center}
\caption{
	Two solid lines represent $\tau$ vs. R1/R2 curves obtained by Monte Carlo 
	simulation for target positions of  $z=0.0$ and 1.0~cm. The hatched zones 
	represent the experimentally deduced R1/R2 ratios for the two target positions. 
	The dashed line denotes the adopted lifetime, and the two dotted lines represent 
	the range of error.}
\label{meanlife}
\end{figure}

	The simulation was then applied to the case of $\gamma$-rays emitted from 
	de-exciting fragments in flight. This simulation employed experimentally obtained 
	parameters such as the energy and emittance of the projectile, the angular spread 
	due to inelastic scattering and multiple scattering, and energy loss of the 
	incoming and outgoing particles. The angular distribution of the $\gamma$-ray 
	emission was also incorporated. The distribution was estimated using ECIS79 
	code~\cite{ecis79}, employing the optical potential parameters determined for 
	the $^{12}$C + $^{12}$C reaction~\cite{OM}. 
	The $\tau$ vs. R1/R2 curves thus obtained for the two target positions are shown 
	in Fig.~\ref{meanlife} as solid lines.

	This figure compares the measured R1/R2 ratios with the simulated curves, 
	allowing the  value of $\tau$ to be determined. The vertical widths of the two 
	hatched zones  represent the ranges of R1/R2 values obtained experimentally. 
	From the overlap between the simulated curves and the hatched zones, the values 
	of $\tau$ were determined as 92~$\pm$~22~ps and 63~$\pm$~17~ps, respectively, 
	for data obtained at target positions of $z=0.0$ and 1.0~cm. The results 
	obtained for the two different target positions are in reasonable agreement. 

	It is probable that the choice of optical potential may affect the calculated 
	$\gamma$-ray angular distribution. This possibility was examined by repeating 
	the distorted wave-born approximation calculation using a different set
	of parameters obtained for the $^{16}$O + $^{12}$C reaction~\cite{OM2}. 
	With the resultant simulation curves, the values of $\tau$ were obtained as 
	93~$\pm$~21~ps and 66~$\pm$~16~ps for the two target positions, $z=0.0$ and 
	1.0~cm. 
	These values are close to those obtained using the original parameters of 
	optical potential. The weighted average of these four values is taken as the 
	final value.

	The uncertainty of $\tau$ considered above, 19\%, is attributable entirely to 
	statistical error. 
	Systematic errors may have arisen primarily from uncertainty 
	of the target position, the accuracy of which was determined to 20\%.
%	$z=0.0$~cm and 20\% for $z=1.0$~cm setting. 
	Another source of systematic error, 
	the 5\% difference in lifetimes determined by the two different sets of optical 
	potential  parameters, should also be taken into account. 
	The total systematic error 
	thus estimated is about 25\%. 
%        Including the statistical and systematic errors,
	By considering all these uncertainties,
	the lifetime was determined to be  $\tau =75 \pm 23$~ps,  
	which corresponds to the \BE\ value of
	$0.63 \pm 0.19 e^2$fm$^4$, or
	$0.26 \pm 0.08$  in Weisskopf units (W.u.).
%	These uncertainties lead to a lifetime of 
%	corresponding to a \BE\ value of 
%	or 

\begin{figure}
\begin{center}
\includegraphics[width=80mm]{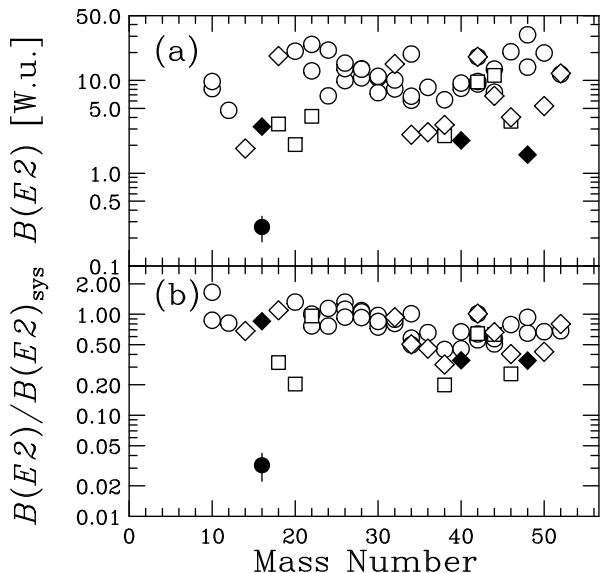}
\end{center}
\caption{
	(a) \BE\ values for $A \le 50$ W.u.  (b) Ratios between \BE\ values and 
	values calculated using the empirical formula~\cite{raman2}. Closed circles 
	denote the values of $^{16}$C, and open circles represent data for other 
	open-shell nuclei. Open squares denote proton closed-shell nuclei, and 
	diamonds indicate neutron closed-shell nuclei. Full diamonds represent double 
	magic nuclei.}
\label{be2all}
\end{figure}

	In Fig.~\ref{be2all}(a), the \BE\ value obtained for $^{16}$C is compared with 
	all the other \BE\ values known for the nuclei with $A<50$~\cite{raman}. Nuclei 
	with open shells tend to have \BE\ values of greater than 10 W.u., whereas 
	nuclei with shell closure of neutrons or protons or both, tend to have distinctly 
	smaller \BE\ values. Typical examples of the latter category are doubly magic 
	nuclei, $^{16}$O and $^{48}$Ca, for which the \BE\ values are  known to be 3.17 
	and 1.58~W.u., respectively. The value of \BE~=~0.26~W.u. obtained for $^{16}$C 
	is even smaller than these extreme cases by as much as an order of magnitude. 

	The anomalously strong hindrance of the $^{16}$C transition can be also 
	illustrated through comparison with an empirical formula based on a liquid-drop 
	model~\cite{raman2}. According to the formula, the values of \E\ and \BE\ are 
	related by   
\[ B(E2)_{\rm sys} = 6.47 \times Z^{2}A^{-0.69}E(2^+_1)^{-1}. \]
	The experimental \BE\ values relative to $B(E2)_{\rm sys.}$ are plotted in 
	Fig.~\ref{be2all}(b). As noted in Ref.~\cite{raman2}, the \BE/$B(E2)_{\rm sys.}$ 
	ratios for open-shell nuclei mostly fall around 1.0, being confined between 0.5 
	and 2.0. Even for the closed-shell nuclei, the ratio remains larger than 0.20. 
	Thus, the ratio of 0.03 for $^{16}$C is exceptionally small, far smaller than 
	for any other nuclei, including closed-shell nuclei.

	Recently a hindered \BE\ has been reported for $^{136}$Te~\cite{Radf}. This 
	nucleus represents 
%	a case analogous to $^{16}$C with respect to neighboring a 
	a case similar to $^{16}$C in the viewpoint of neighboring a 
	doubly magic nucleus. 
	The \BE\ value observed for $^{136}$Te is 206~$e^2$fm$^4$, which corresponds 
	to \BE/$B(E2)_{\rm sys.}=0.21$. Thus, the $E2$ strength is indeed hindered. 
	However, the degree of the hindrance is comparable to that of singly closed 
	nuclei, and is far more moderate than for the case of  $^{16}$C.

	In considering the nature of the $^{16}$C transition, it is instructive to 
	invoke a simple seniority-two empirical model~\cite{Radf} to describe the wave 
	function of the 2$_1^+$ state of $^{16}$C. The nucleus $^{16}$C has two missing 
	protons and two extra neutrons with respect to $^{16}$O. Then, the mixing of 
	$\pi^{-2}$ and $\nu^2$ configurations with a residual interaction of $V$ can 
	be considered by taking the observed 2$_1^+$ states of $^{14}$C and $^{18}$O 
	as the basis states. The two singly closed isotopes, $^{14}$C with $N=8$ 
	and $^{18}$O with $Z=8$, have two missing protons and two extra neutrons 
	with respect to $^{16}$O, respectively. Thus, the  2$_1^+$ state of $^{14}$C 
	may be regarded as primarily due to proton excitations, while that of $^{18}$O 
	can be attributed almost entirely to neutron excitations. The $E(2_1^+)$ values 
	are 7.01 and 1.98~MeV, and the \BE\ values are 3.70 and 9.53~$e^2$fm$^4$ 
	(or 1.84 and 3.40~W.u.) for $^{14}$C and $^{18}$O, respectively. 
	The very large $E(2_1^+)$ for $^{14}$C indicates a large energy gap between 
	the two proton orbitals, 1$p_{3/2}$ and 1$p_{1/2}$, while the moderate energy 
	for $^{18}$O may reflect unoccupied $sd$-shell orbitals available for neutron 
	excitation. Correspondingly, \BE\ for $^{14}$C is close to 1~W.u., whereas 
	that for $^{18}$O is somewhat larger. 

	By applying the empirical model, $|V|=1067$~keV is required to generate the 
	1766-keV state in $^{16}$C. The very high energy of $|\pi^{-2}\rangle$ results 
	in a configuration dominated by $|\nu^2\rangle$: 
	\[|2^+_1;^{16}{\rm C}\rangle = 0.20 |\pi^{-2}\rangle - 0.98 |\nu^2 \rangle. \]
	Here, the interaction $V$ between proton-holes and neutron-particles is 
	repulsive in the present case~\cite{hama}, leading to destructive interference 
	between the two $E2$ components. The calculated \BE\ for $^{16}$C is 
	7.0~$e^2$fm$^4$, which falls between these components 
	of $^{14}$C and $^{18}$O, and 
	overestimates the experimental \BE\ value of $^{16}$C by almost an order of 
	magnitude. This is in contrast to the case for $^{136}$Te, which when treated 
	by seniority-two scheme provides a fairly reasonable estimate of the experimental 
	\BE~\cite{Radf}. The unprecedented hindrance for $^{16}$C is suggestive of an 
	anomalous nuclear structure.

	In this regard, several unique features of the nuclei in the vicinity of 
	$^{16}$C can be noted. For example, it has been suggested~\cite{fuji} that 
	the energy gaps between single-particle orbitals such as 1$p_{3/2}$ and 
	1$p_{1/2}$, may vary appreciably as the isotope changes from $^{14}$C to 
	$^{16}$C. It has also been indicated based on electric quadrupole moment 
	measurement of neighboring nuclei $^{15,17}$B~\cite{15B,17B} that 
	the $E2$ effective charges can be markedly reduced for extremely neutron-rich 
	nuclei. 
	Furthermore, delicate interplay between clustering and deformation
	may prevail in these light nuclei~\cite{amd}. Elaborate calculations involving 
	such exotic features can be expected to be helpful in accounting for the 
	hindrance of \BE\ observed for $^{16}$C. 

	The authors would like to thank the RIKEN Ring Cyclotron staff for cooperation 
	during the experiment. N.~I. is grateful for the financial assistance from 
	the Special Postdoctoral Researcher Program of RIKEN. The present work was 
	supported in part by a Grant-in-Aid for Scientific Research (No.~1520417) from 
	Monbukagakusho (Japan) and OTKA 042733 (Hungary).

\end{document}